\documentstyle[aps,twocolumn,prb,epsfig,epsf]{revtex}

\def\bk{{\bf k}}
\def\bq{{\bf q}}
\def\bkp{{\bf k}'}

\def\bgeq{\begin{equation}}
\def\eneq{\end{equation}}
\def\bgeam{\begin{mathletters}\begin{eqnarray}}
\def\eneam{\end{eqnarray}\end{mathletters}}
\def\bgea{\begin{eqnarray}}
\def\enea{\end{eqnarray}}

\begin{document}
\draft
\title{Sign-reversal of drag in bilayer systems with in-plane periodic potential modulation}

\author{Audrius Alkauskas,$^{1,2}$ Karsten Flensberg,$^1$ Ben Yu-Kuang Hu,$^3$
Antti-Pekka Jauho,$^4$}

\address{$^1$ {\O}rsted Laboratoriet, Universitetsparken 5, K{\o}benhavns Universitet,
DK-2100 K{\o}benhavn \O, Denmark\\
$^2$ Physics Faculty, Vilnius University, Saul\.{e}tekio 9,
Vilnius 2040, Lithuania\\
$^3$ Department of Physics, University of Akron, Akron OH 44325-4001
$^4$ Mikroelektronik Centret, Danmarks Tekniske Universitet,
DK-2800 Lyngby, Denmark
\\\medskip
\date{\today}
\parbox{14cm}{\rm
We develop a theory for describing frictional drag in bilayer
systems with in-plane periodic potential modulations, and use it
to investigate the drag between bilayer systems in which one of
the layers is modulated in one direction. At low temperatures, as
the density of carriers in the modulated layer is changed, we show
that the transresistivity component in the direction of modulation
can {\it change its sign}.  We also give a physical explanation
for this behavior.}
\smallskip\\}
\maketitle
\narrowtext
Throughout past decade there has been a great deal of experimental
and theoretical activity in frictional drag in bilayer systems,
following the seminal experiments by Gramila {\em et
al.}\cite{gram91}  These drag experiments involve a double quantum
well system where the layers are individually contacted by
ingenious fabrication techniques.  The barrier between the wells
is made thick enough to suppress tunneling but thin enough to
allow significant interlayer interactions. An average current
density ${\bf j}_1$ is driven through layer 1 and circuit is kept
open in layer 2, so that ${\bf j}_2 = 0$.  The interlayer
interaction causes the electrons in layer 1 to drag along the
electrons in layer 2, and hence a counterbalancing electric field
${\bf E}_2$ forms in layer 2 to maintain a zero net ${\bf j}_2$.
The transresistivity tensor $\tensor{\rho}_{21}$, defined by ${\bf
E}_2 = \tensor\rho_{21} {\bf j}_1$, can be extracted
experimentally and can reveal important information about the
properties of the effective interlayer interactions, the
individual layers and the coupled bilayer system.

Since Gramila {\em et al.}'s original work\cite{gram91}, which was
done on a closely spaced electron--electron system at low
temperatures without an applied magnetic field, many variations on
the theme of the original experiments have been performed.  For
instance, drag has been measured in electron--hole\cite{siva92}
and hole--hole\cite{jorg00} systems, in widely separated
layers\cite{gram93,gram94}, and in the presence of a perpendicular
magnetic field.\cite{jorg00,hill96,rube97,lill98}  Very recently,
low--density systems have been studied to probe the suggested
metal--insulator phase transition in strongly correlated
disordered two-dimensional systems.\cite{pilla02,hwan02} Another
modern trend is to examine mesoscopic effects in Coulomb
drag.\cite{naro00,mort01,mort02}. In general, drag without an
applied $B$-field is reasonably well understood within the
framework of a standard weak-interlayer coupling
theory.\cite{jauh93,kame95,flen95a} The theory successfully
accounts for several unusual features such as large enhancements
in the transresistivity (up to an order the magnitude; some
intriguing discrepancies however do persist for the most dilute
systems studied\cite{hwan02}) due to intralayer
correlations\cite{swie95} and plasmon mediated
scattering.\cite{flen94,flen95b} On the other hand, the
understanding of magnetodrag ({\em i.e.}, drag in the presence of
a perpendicular $B$-field) in bilayer systems is less complete,
and several puzzling experimental results remain unexplained.  For
instance, under certain circumstances, the diagonal terms in the
magnetotransresistivity ($\rho_{21}^{xx}$ and $\rho_{21}^{yy}$)
has been observed to {\em reverse sign} when the chemical
potential is changed in one layer while being kept fixed in the
other.\cite{lok01} This sign reversal with changing chemical
potential (which incidentally has not been observed at $B=0$)
cannot be obtained from magnetodrag calculations using the
self-consistent Born approximation,\cite{bons96,bons97} and
despite recent theoretical progress,\cite{oppe01} a fully
satisfactory explanation of this phenomenon is not yet available.

In this paper, we suggest that a reversal of the sign of the
transresistivity is possible at $B=0$ in bilayer systems that have
periodic potential modulations in the plane of the layers. The
periodic potential modulation creates mini-bands, and the charge
carriers can evolve from electron-like to hole-like behavior with
a relatively small change in the density. Furthermore, for systems
that are modulated in one direction, it is possible to observe
``skewed drag"  ({\em i.e.}, non-zero off-diagonal elements of
$\tensor{\rho}_{21}$), implying that the electric field response
in the drag layer is in a different direction from that of the
driving current.  This demonstrates the important role
band-structure plays in determining the transresistivity of the
system.\cite{albr99}  We note that experiments on two-dimensional
electron gases with strong potential modulations in one direction
have already been reported in the literature\cite{deut01}, and
hence we believe that the theory described below is amenable to
experimental tests in near future.

To investigate drag in these modulated systems, we use the Kubo
formalism\cite{kame95,flen95a} to calculate the transconductivity
tensor $\tensor\sigma_{21}$, which is related to the
transresistivity by $\tensor\rho_{21} =
-\tensor\rho_{22}\tensor\sigma_{21}\tensor\rho_{11}$ in the weak
interlayer coupling limit. In this method, the transconductivity
is expressed as a current-current correlation function, which can
be calculated with standard perturbation theory techniques.

 The Hamiltonian of the system is $\hat{H} =
\sum_{i=1,2}\hat{H}_{i} + \hat{H}_{12},$ where $\hat{H}_i$ is the
Hamiltonian of layer $i$ and $\hat{H}_{12}$ is the interlayer
interaction term.  We assume $\hat{H}_{21}$ is due to Coulomb
interactions, so that $ \hat{H}_{12} = {\cal A}^{-1}\sum
\hat{n}_1({\bf q}) \hat{n}_2(-{\bf q}) V_{12}({\bf q}),$ where
$\hat{n}_i({\bf q})$ and $V_{12}({\bf q})$ are the Fourier
transforms of the density operator and the interlayer Coulomb
interaction, respectively.

We define, within the Matsubara formalism, $\vec\Delta$ to be the
correlation function\cite{differentDelta}
\begin{eqnarray}
\vec\Delta({\bf q},{\bf q}';i\omega_n,i\omega_n') &=&
-\int_0^\beta d\tau \int_0^\beta d\tau'e^{i\omega_n\tau}
e^{-i\omega_n'\tau'}\nonumber\\
&\times& \langle T_\tau \hat{\bf J}(0)\; \hat{n}({\bf q},\tau)
\;\hat{n}(-{\bf q}',\tau') \rangle_0 .
\end{eqnarray}
For systems that have a periodic potential modulation with
reciprocal vectors {\bf G}, only ${\bf q} - {\bf q}' = {\bf G}$
terms are non-zero. Expanding in powers of $V_{12}$, the first
non-vanishing term for $\sigma_{21}$ in the dc limit is the second
order term. We obtain
\begin{eqnarray}
\sigma_{21}^{\delta\gamma} &=& \frac{e^2}{h{\cal A}} \sum_{\bf q}
\sum_{\bf G_1 G_2} V_{12}({\bf q}) V_{12}(-{\bf q}+{\bf
G_1})\delta_{\bf G_1G_2}\nonumber\\
&\times& \int_0^\infty \frac{d\omega}{2\pi}  \Delta_2^\delta({\bf
q},{\bf q}+{\bf G_2};\omega
+i0^+,\omega-i0^+)\nonumber\\
&\times& \Delta_1^\gamma({\bf q},{\bf q}+{\bf
G_1};\omega+i0^+,\omega-i0^+) [-\partial_\omega
n_B(\omega)]
\label{sigma21}
\end{eqnarray}

Evaluation of $\vec\Delta({\bf q},{\bf q}+{\bf
G};\omega+i0^+,\omega-i0^+)$ is analogous to Ref.
\onlinecite{flen95a}. In this paper, we assume the system is in
the weak scattering limit, allowing us to ignore vertex
corrections at the charge vertices. Then, one obtains
\begin{eqnarray}
&&\vec\Delta({\bf q},{\bf q}+{\bf G},\omega
+i0^+,\omega-i0^+)\nonumber\\
&& =\frac{4\pi}{\cal A}\sum_{\bk nn'} \left[{\bf v}_{n'{\bf
k+q}}\tau_{{\rm tr},n'}({\bf k+q}) - {\bf v}_{n{\bf k}}\tau_{{\rm
tr},n}({\bf k}) \right]
\nonumber \\
&&\times \left[( n_{F}(\varepsilon _{n{\bf k}})-n_{F}(\varepsilon _{n{\bf k}%
}-\omega )\right] \delta \left( \varepsilon _{n{\bf
k}}-\varepsilon _{n^{\prime }{\bf k+q}}-\omega \right)
\nonumber\\
&&\times\eta(\bk+\bq n', \bk n; \bq)\; \eta(\bk n, \bk+\bq n'; -\bq - {\bf G}).
\label{deltaballistic}
\end{eqnarray}
Here, ${\bf v}_{n{\bf k}}$ is the band velocity, $\varepsilon$ is
the energy,
 $n_F(\varepsilon) = [\exp(\beta(\varepsilon-\mu))+1]^{-1}$
($\mu$ is the chemical potential), $\tau_{\rm tr}$ is the
transport time, $\eta({\bf k}'\, n',{\bf k} n;{\bf q}) = \langle
{\bf k}'\, n'|\exp(-i{\bf q}\cdot{\bf r})|{\bf k} n\rangle$
($n,n'$ are the band indices)]. The $\tensor\sigma_{21}$ obtained
using Eq.~(\ref{sigma21}) and the weak scattering result,
Eq.~(\ref{deltaballistic}), can also be derived from the
semiclassical Boltzmann equation.\cite{flen95b}

A complete calculation of drag, using (\ref{deltaballistic}) in
(\ref{sigma21}), is an arduous task, requiring a numerical
evaluation of the band-structure(s) of the layers, calculation of
the matrix elements $\eta(\bkp n', \bk n; \bq)$, and summation of
different bands $n$, and reciprocal lattice vectors $\bf G$. For
incommensurate lattices one always has ${\bf G_1}=0={\bf G_2}$,
and in the remaining part of the paper we assume this to be the
case. The other technical steps do not pose conceptual
difficulties, and in the present context we find it  appropriate
to consider simplified systems where to a certain extent analytic
progress can be made, and for which the physics is transparent.

The central issue of this paper is the sign reversal of the drag.
We demonstrate this first for a 1--dimensional model, neglecting
interband processes and the momentum dependence of the transport
relaxation time.  For this case, the correlation function
(\ref{deltaballistic}) becomes $\Delta=\tau_{\rm tr} F(q,\omega)$,
where
\begin{eqnarray}
F(q,\omega)&=&-\frac{2\pi}{\cal
L}\sum_k\left(v_k-v_{k+q}\right)\nonumber\\
&\quad&\times\left[n_F(\varepsilon_k)-n_F(\varepsilon_{k+q})\right]
\delta\left(\varepsilon_k-\varepsilon_{k+q}-\omega\right)\nonumber\\
&=&\sum_{k_i} {\rm sign}\left(v_{k_i}-v_{k_i+q}\right)
\left[n_F(\varepsilon_{k_i})-n_F(\varepsilon_{k_i+q})\right],
\end{eqnarray}
where $k_i$ are the solutions of
$\varepsilon_{k_i}-\varepsilon_{k_i+q}-\omega=0$.  For
illustrative purposes, we consider a cosine-band, $\varepsilon_k=
-\hbar^2/(ma^2) \cos{ka}$, for which there are two (or none)
solutions, and one finds
\begin{eqnarray}
F(q,\omega)&=&{\rm
sign}\left(v_2-v_1\right)\left[n_F(-\varepsilon_2)-n_F(\varepsilon_1)
\right.\nonumber\\
&\quad&\quad\left.-n_F(-\varepsilon_1)-n_F(\varepsilon_2)\right].
\label{1dtranspol}
\end{eqnarray}
At half--filling the chemical potential $\mu$ vanishes, and making
use of $n_{F,\mu=0}(-\varepsilon)=1-n_{F,\mu=0}(\varepsilon)$, it
is easy to see that the result (\ref{1dtranspol}) vanishes
identically.  Thus, in an experiment where one of the subsystems
is kept unchanged while in the other the chemical potential is
moved through half--filling, the drag will change sign.  While the
above discussion is an important demonstration of principle, it is
necessary to also consider periodically modulated two-dimensional
electron gases, which are the most commonly studied systems in
this context.\cite{gudi02}

A system which has an identical periodic modulation in both $x$--
and $y$--directions is characterized by particle--hole symmetry,
and it seems natural that the drag passes through zero when the
two carrier species are matched.  The experimentally most relevant
systems are those, however, where the modulation is only in one
direction\cite{deut01,weis89} (the strongest modulations have been
achieved for these systems) and hence we choose the model system
as follows: (1) There is a single band (the dispersion law and
corresponding density of states are illustrated in Fig. 1) with a
tight-binding dispersion relation $\varepsilon(k_x,k_y) =
\hbar^2[1 - \cos(k_x a)]/(m_x a^2) + \hbar^2 k_y^2/(2 m_y), $ and
hence the velocity components are $ v_x(k_x) = \hbar \sin(k_x
a)/(m_x a) $ and $v_y(k_y) = \hbar k_y/m_y.$ (2) $\tau_{\rm tr}$
is {\bf k}-independent. (3) The interlayer interaction $V_{12}(q)$
is significant only for small $q$.\cite{v(q)}

At low temperatures, it would appear permissible to expand in
$\omega$, because $\partial n_B(\omega)/\partial\omega$ in the
integrand cuts off the higher $\omega$ contributions.  Following
this procedure yields an analytic expression for $F$. and the
resulting drag resistivity obeys the familiar $T^2$--dependence
known from unmodulated two--dimensional
systems\cite{gram91,jauh93}. In this scheme, the transresisitivity
{\it diverges} when $\mu= 2\hbar^2/(m_xa^2)$ with the opposite
sign from the low density $\rho^{xx}$ . This divergence, which is
related to the divergence in the density of states at this energy
(see Fig. 1), is unphysical because it only occurs in the
experimentally unreachable $T=0$ limit. Nevertheless, it is
interesting to observe that
this approximation leads to a change of sign in the
transconductance, and we should expect this behavior to be most
prominent when $\mu= 2\hbar^2/(m_x a^2)$.

To cure this spurious divergence one must avoid expansions, and
perform a numerical evaluation.  As a starting point we have found
it convenient to use
\begin{eqnarray}
F^x\left( \bq,\omega \right)  &=&\frac{m_{y}}{\left( 2\pi \right)
^{2}\hbar q_{y}m_{x}a}
\int\limits_{-\pi/a+q_{x}/2}^{\pi/a-q_{x}/2}
dk_{x}\nonumber\\
&\times&\left[ \sin \left( k_{x}-q_{x}/2
\right) a-\sin \left( k_{x}+q_{x}/2\right) a\right]\nonumber  \\
&&\times \left[
n_F\left(k_{x}-q_{x}/2,k_{y0}\left(k_{x}\right)-q_{y}/2\right)\right.\nonumber\\
&-&\left. n_F\left( k_{x}+q_{x}/2, k_{y0}\left(k_{x}\right)
+q_{y}/2\right) \right] , \label{Fright}
\end{eqnarray}
where
\begin{eqnarray}
k_{y0}\left( k_{x}\right) &=&\frac{m_{y}}{\hbar^{2}q_{y}}
\bigl[\hbar \omega + \frac{\hbar ^{2}}{m_{x}a^{2}}(\cos
\left( k_{x}+q_{x}/2\right) a\nonumber\\
&\quad&-\cos \left( k_{x}-q_{x}/2\right) a) \bigr] .
\end{eqnarray}

While the Kubo formula gives the transconductivity, it is often
most convenient to express the results in terms of
transresisistivity (this is the object usually recorded in
experiments) $\tensor\rho_{21}$, whose components are given by
\begin{equation}
\rho_{21}^{xx}=-\frac{\sigma^{xx}_{21}}{\sigma_{11}^{xx}\sigma_{22}^{xx}
-\sigma_{12}^{xx}\sigma_{21}^{xx}}\simeq
-\frac{\sigma^{xx}_{21}}{\sigma_{11}^{xx}\sigma_{22}^{xx}},
\end{equation}
and analogously for the $yy$-component.  The transresisivity
tensor has the additional advantage that is does not involve the
transport relaxation times for the individual layers, as long as
these are momentum independent.  The computed transresisitivity is
shown in Fig. 2 for four different temperatures. The most
important feature is that the drag indeed changes sign; the effect
is most prominent for low temperatures, and densities close to a
fully-occupied band.  In Figs. 3 and 4 we compare the $xx$- and
$yy$-components of the computed transconductances (the
$xx$-component was used in calculating the results of Fig. 2). We
observe that the sign change does not take place for
$\sigma_{12}^{yy}$, nevertheless an interesting double-peak
structure emerges.

An analysis of the several assumptions made in our calculations is
now in place. We have assumed that the system only has one band.
Clearly, this assumption breaks down when the density so  large
that Fermi energy significantly exceeds $2\hbar^2/(2m_xa^2)$,
because the carriers will start to occupy higher bands. We also
have assumed that temperature is low enough that the inelastic
mean free path $\ell_{\rm in}$ is much longer than period of the
potential modulation, $a$.  For finite $\ell_{\rm in}$ the system
is roughly divided into coherent regions of order $\ell_{\rm
in}^2$.  If $\ell_{\rm in} \lesssim a$, the electrons do not
coherently feel the periodic potential, and the drag
characteristics will be given by an average of the drag over the
density fluctuation caused by the potential modulation.  Since the
system as a whole acts like a (nearly) uniform system in this
case, effects described in this paper will not be observable at
temperatures for which $\ell_{\rm in} \lesssim a$.

To summarize, we have  developed a theory for drag in bilayer
systems where there is a periodic potential modulation.  We have
calculated the drag for the case where there is potential
modulation in one direction in one of the two layers. We find that
at low temperatures the transresistivity changes sign as the
density is increased.  The anisotropy of the transresistivity
tensor implies that one should be able to see Hall drag.
Experimentally, it may be possible to fabricate the system
investigated here by by overgrowing a pair of quantum wells over a
cleaved edge,\cite{deut01} or using lithographic
techniques.\cite{gram-priv}

AA acknowledges a NORFA grant which was crucial for this study,
and BYKH is supported by DOE Grant No.\ DE-G02001ER45948, Research
Corporation Award No.\ CC5168, and was supported by a University
of Akron Summer Research Fellowship.

\begin{figure}
\begin{center}
\epsfig{file=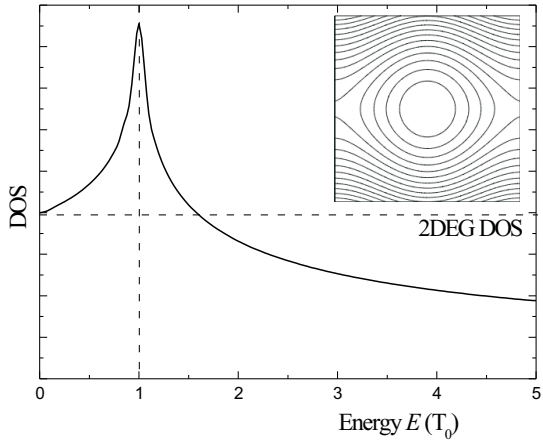, width=0.9\columnwidth,clip}
\end{center}
\caption{The density of states for a two-dimensional system with
periodical modulation in one direction.  The inset shows the
constant energy surfaces for the dispersion relation used in this
work.  The energy is in units of $k_BT_0=2\hbar^2/(m_xa^2)$.}
\end{figure}

\begin{figure}
\begin{center}
\epsfig{file=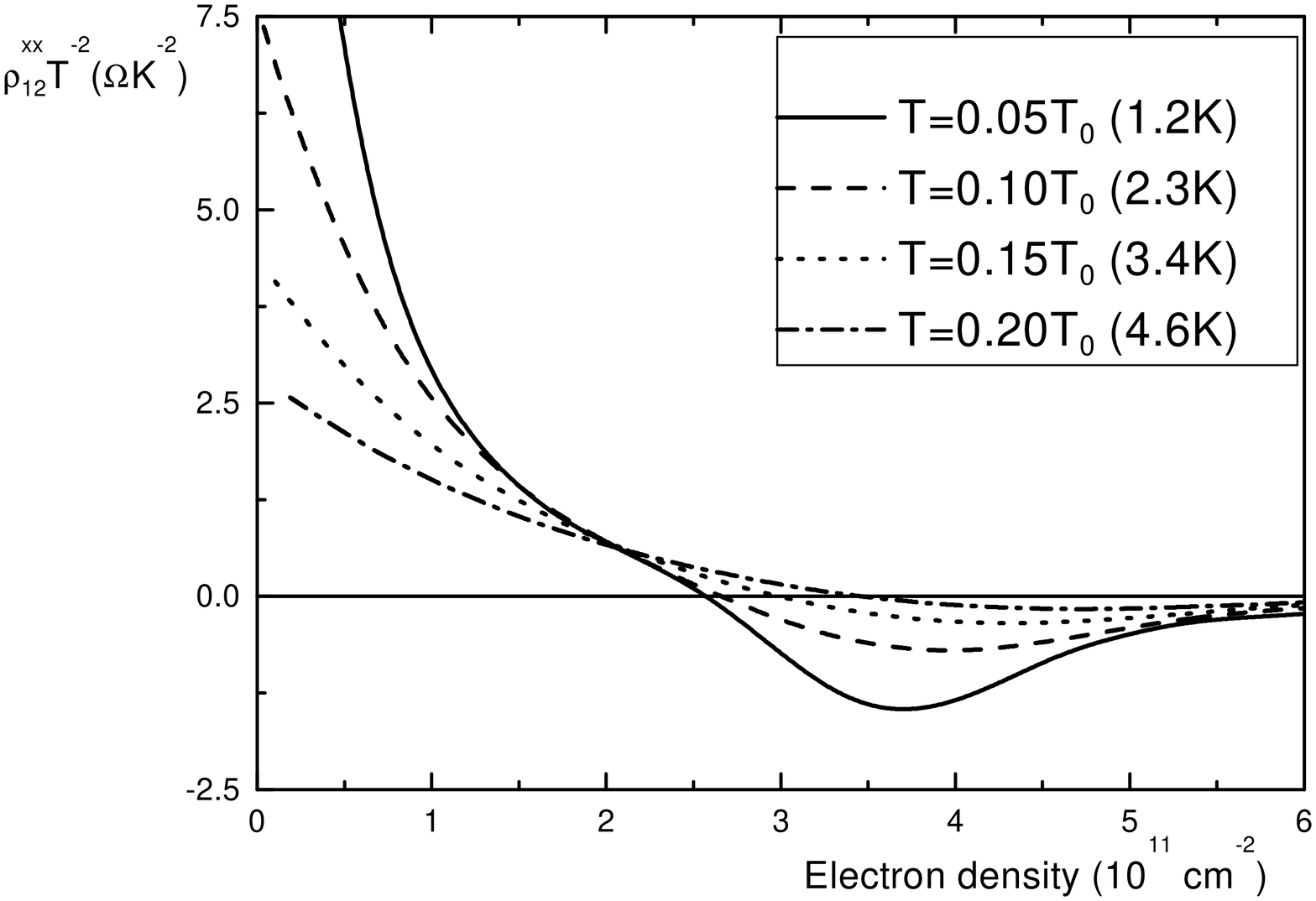, width=0.9\columnwidth,clip}
\end{center}
\caption{The calculated transresistivity $\rho^{xx}_D$, as a
function of the density for four different temperatures.}
\end{figure}

\begin{figure}
\begin{center}
\epsfig{file=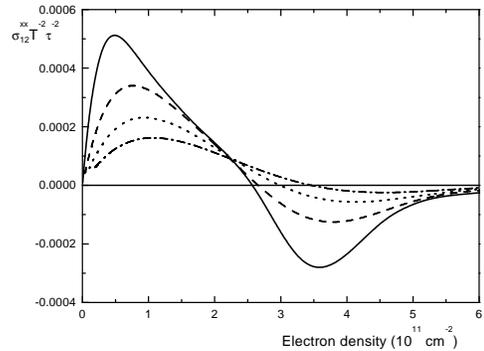, width=0.9\columnwidth,clip}
\end{center}
\caption{The normalized transconductivity $\sigma^{xx}_D$, as a
function of the density for the same temperatures as in Fig. 1. }
\end{figure}

\begin{figure}
\begin{center}
\epsfig{file=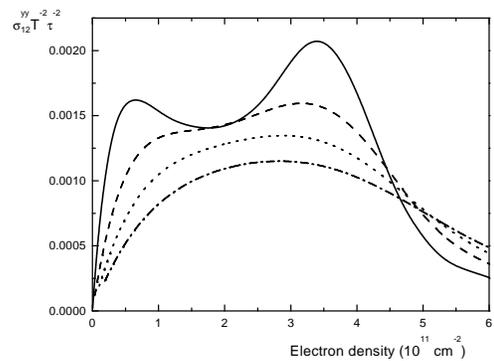, width=0.9\columnwidth,clip}
\end{center}
\caption{The normalized transconductivity $\sigma^{yy}_D$, as a
function of the density for four different temperatures.}
\end{figure}


\begin{thebibliography}{99}
\bibitem{differentDelta}  Note that this definition is different from the
$\Delta$ used in Ref.~\onlinecite{flen95a}.  The $\Delta$ defined
in Ref.~\ \onlinecite{flen95a} is appropiate only for homogenous
systems.
\bibitem{v(q)} In our calculations we use the statically screened
Coulomb interaction, $V(q)=(2\pi e^2/\kappa)q/[2q_{{\rm
TF},1}q_{{\rm TF},2}\sinh{qd}+q(q+q_{{\rm TF},1}+q_{{\rm
TF},2})\exp{qd}]$, where $q_{{\rm TF},i}$ are the Thomas-Fermi
screening wave-numbers for the two layers.  For the modulated
layer the screening wave-number has a slight temperature and
density dependence, which, however, has a rather small effect for
the calculated drag.
\end{thebibliography}
\end{document}